\documentclass{optica-article}

\journal{opticajournal} 

\articletype{Research Article}

\usepackage{graphicx}
\usepackage{grffile}
\usepackage{array}
\usepackage{varwidth} 
\usepackage{nameref} 
\usepackage{ulem} 

\usepackage{textcomp}

\usepackage{lineno}
\newcolumntype{V}[1]{>{\centering\arraybackslash}m{#1}}
\begin{document}

\title{Eikonal phase retrieval: Unleashing the fourth generation sources potential for enhanced propagation based tomography on biological samples.}

\author{
  Alessandro Mirone,\authormark{1,*}
  Joseph Brunet,\authormark{1,2}
  Leandre Admans,\authormark{1}
  Renaud Boistel,\authormark{3}
  Morgane Sowinski,\authormark{3,4}
  Camille Beruyer,\authormark{1,2}
  Henry Payno,\authormark{1}
  Elodie Boller,\authormark{1}
  Pierre Paleo,\authormark{1}
  Claire L. Walsh,\authormark{2}
  Peter D. Lee,\authormark{2} and
  Paul Tafforeau\authormark{1}
}

\address{\authormark{1}European Synchrotron Radiation Facility, 71, Avenue des Martyrs, Grenoble F-38000, FR\\
\authormark{2}Department of Mechanical Engineering, University College London, London, UK.\\
\authormark{3} UMR 7179 Mus\'eum National d' Histoire Naturelle, 57, rue Cuvier, CP 55, 75005 Paris, FR.
\authormark{4}Laboratoire d' Acoustique de l'Universit\'e du Mans (LAUM), UMR 6613, Institut d'Acoustique-Graduate School (IA-GS), CNRS, Le Mans Universit\'e, Avenue Olivier Messiaen 72085, Le Mans, FR.

}

\email{\authormark{*}mirone@esrf.fr} 


\begin{abstract*} 
  The evolution of synchrotrons towards higher brilliance beams has increased the possible sample-to-detector propagation distances for which the source confusion circle does not lead to geometrical blurring. This makes it possible to push near-field  propagation driven phase contrast enhancement to the limit, revealing low contrast features which would otherwise remain hidden under an excessive noise. Until today this possibility was hindered, in many objects of scientific interest, by the simultaneous presence of strong phase gradient regions and low contrast features. The strong gradients, when enhanced with the now possible long propagation distances, induce such strong phase effects that the linearisation assumptions of current state-of-the-art single-distance phase retrieval filters are broken, and the resulting image quality is jeopardized. Here, we introduce a new iterative phase-retrieval algorithm and
compare it with the Paganin phase-retrieval algorithm, in both the monochromatic and polychromatic cases, obtaining superior image quality. 
 In the polychromatic case the comparison was done with an extrapolated Paganin algorithm obtained by reintroducing, into our phase retrieval algorithm, the linearization approximations underlying the Paganin forward model. Our work provides an innovative algorithm which efficiently performs the phase retrieval task over the entire near-field range, producing images of exceptional quality for mixed attenuation objects.
\end{abstract*}

\section{Introduction}
Among the phase contrast techniques employed at synchrotrons, Propagation Phase-Contrast Micro-Tomography (PPC-$\mu$CT) is the most commonly used, both in term of devoted beamlines and imaging throughput. This is due to its simplicity, and to its photon cost efficiency, with a PPC-$\mu$CT implementation consisting of a simple detector downstream-translation applied to a standard radiography setup. The usage of the large bandwidth beam generated by a wiggler or a bending magnet, with only filters, but no crystal or multilayer monochromator, is appropriate for propagation phase contrast techniques \cite{thenewbm18}.
In contrast, far-field techniques require more sharply monochromatized beams with longer temporal (also known as longitudinal) coherence. This, however, comes at the cost of reduced photon flux and typically requires either several distances of propagation, or displacement of masks in the beam. In these techniques, speckles and fringes result from the interference of radiation originating from distant parts of the wavefront, and consequently, different parts of the sample, following paths of varying lengths \cite{id16a}.
In PPC-$\mu$CT, most of the interference fringes disappear, smeared out by the short temporal coherence, but the effect of the transversal derivatives of the wavefront phase remains. These phase derivatives are induced by the phase density gradients of the sample. The derivative being a local operator, when acting at a given point of the wavefront, the transversal phase derivatives effects remain with short longitudinal coherence \cite{Pogany1997}. 
A familiar example in the visible domain would be the heat shimmers seen above a sunbathed hot asphalt road. Such shimmers result from propagation-enhanced phase-contrast induced by the vertical density gradient of the air. They become visible when the effects of the small deviations of the light are amplified by a long propagation. In the X-ray domain the  ratio between the imaginary part and the decrement of the refraction index is positive \cite{xraybooklet}. Consequently, the focusing (defocusing) effect of the phase second derivatives, the lensing effect, intensifies the brightness (darkness) of a transparent (absorbing) feature. The visibility of small phase gradients increases with propagation distance as long as their induced angular deviation is larger than both source and detector pixel angular diameters as seen from the sample. Thanks to this increased visibility, propagation phase-contrast imaging enhances the signal above Poisson noise without increasing the absorbed dose, which is often detrimental to the samples. Moreover, it also increases the signal ratio over systematic errors, such as those due to irregularities in the detection devices, or inhomogeneities in the beam profile. Simply extending the counting time in absorption-based imaging wouldn't suffice to reduce these systematic effects. The formalism that is applied in the analysis of propagation phase-contrast radiographs aims at deducing the intensity distribution of a fictive wavefront plane situated immediately after the sample, given the detector measured intensity. The latter contains strong phase effects, with the propagation enhanced signal strengthened over the statistical noise and systematic errors, while the retrieved intensity, deduced from such phase enhanced signal, is simply what one could measure based on the sample absorption effect only, and has reduced noise. Once the intensity at the fictive wavefront plane is known, the reconstruction problem is reduced to the one of standard absorption tomography, but with an improved signal over noise ratio compared to what one would measure experimentally without the long propagation distance. The simplicity of the experimental setup, combined with the drastic improvement of the signal-to-noise ratio, explains the wide usage of propagation phase-contrast techniques \cite{pcxmedical}.
One key application area of tomography at synchrotron sources is the imaging of biological samples with about one third of all the accepted proposals at the three PPC-$\mu$CT  ESRF beamlines BM18, ID19, BM05, over the last year, for a total of 2032 hours of beamtime \cite{useroffice}. One of the most fascinating properties of biological samples is that their structural hierarchy is ubiquitous at all scales. This implies that the signal contains both weak and strong components. For instance, a sample containing tiny blood vessels, which are visible thanks to the weak refractive index differences between tissues and liquids, might also contain sharp edges in the optical index at the interfaces of the bone structures. Increasing the propagation distance may enhance visibility of the weak components, but this also increases the signal from the strong components. We demonstrate that this enhancement may exceed the validity limits of current algorithms, generating streak artifacts in the reconstruction.
In this work, we have extended the theory underlying PPC-$\mu$CT algorithms by improving the reconstruction quality in the case of strong phase gradients and large dynamics in the transmitted intensities (see Methods for details). The algorithm is based on the eikonal approximation. The term {\it eikonal} made its first appearance, in physics, in studies of wavelike phenomena (waves in channels, optics, and then quantum mechanics \cite{landau}), where the amplitude of a wavefront is calculated by integrals along the minimum action paths, those which in geometrical optics form the images alias {\it icons}, hence the Greek root name. We use the same approach to formalise our method. Accordingly, it has been named the Eikonal Phase Retrieval (EPR) method. Moreover, as our studies bring us in the high contrast regime, as detailed in Methods, we also introduce an analytical ansatz for the scintillator point spread function. Using this ansatz in the data preprocessing phase, by deconvolution \cite{deconv}, we efficiently remove the low frequency artifacts caused by the non-uniformity of the X-ray illumination on the scintillator. This latter algorithm accounts for the deconvolution of the scintillator diffused secondary light background and is called Scintillator Light Deconvolution (SLD).

\section{Results}

The EPR and SLD methods were validated on two biological samples: a sheep head (purchased in a butchery and then skinned), and a giant frog of the species Leptodactylus pentadactylus (smoky jungle frog, collection number IRSNB 391F). Both specimens were preserved in 70\% ethanol and mounted following the protocol for Hierarchical Phase-Contrast Tomography (HiP-CT) \cite{Brunet2022}.  They were then scanned using the HiP-CT method \cite{Walsh2021}. The sheep head presents a challenge due to its strong absorption contrast, which arises from long paths inside bones and the strong phase gradients at both their external surfaces and inner structures. In contrast, the frog has lower absorption and tinier features. Both samples are representative of the research conducted at BM18, in particular in the frame of the Human Organ Atlas Project \cite{cziproject}.

We performed the tomography of the sheep head at an average BM18 beam energy of 109keV using a pixel size of 28$\mu$m and a propagation distance of 30 meters.

\begin{figure}[htbp]                                                                                                                                                                                       
\centering\includegraphics[width=\textwidth ]{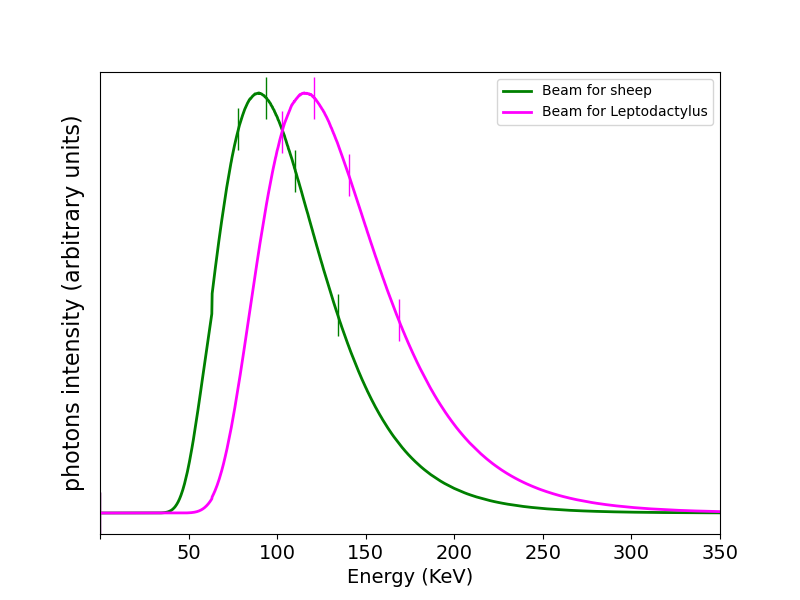}                                                                                                   
  \caption{Spectral shape used for the two HiP-CT scans. Green curve: sheep head experiment; Magenta curve: {\it Leptodactylus pentadactylus} experiment. The vertical lines delimit the intervals used in the discretisation of the spectral shape.  }
  \label{spectramouton}
\end{figure}

Figure \ref{spectramouton} shows the relevant X-ray spectra. In the HiP-CT method the radiographs are taken of a sample immersed in a cylindrical jar, larger than the field of view, filled with ethanol, and the sample signal is obtained dividing the measured intensities by the radiographs of the same jar filled with ethanol only. The figure shows the beam effective spectral shape after conversion by the scintillator and detector signal collection, taking into account the absorption paths and conversion and detection efficiency. The spectra reaching the sample depends on the wiggler parameters and the filters which are used to attenuate the beam and change its spectral shape \cite{Tanaka2001,xop}.

\begin{figure}[htbp]                                                                                                                                                                                       
\centering\includegraphics[width=\textwidth ]{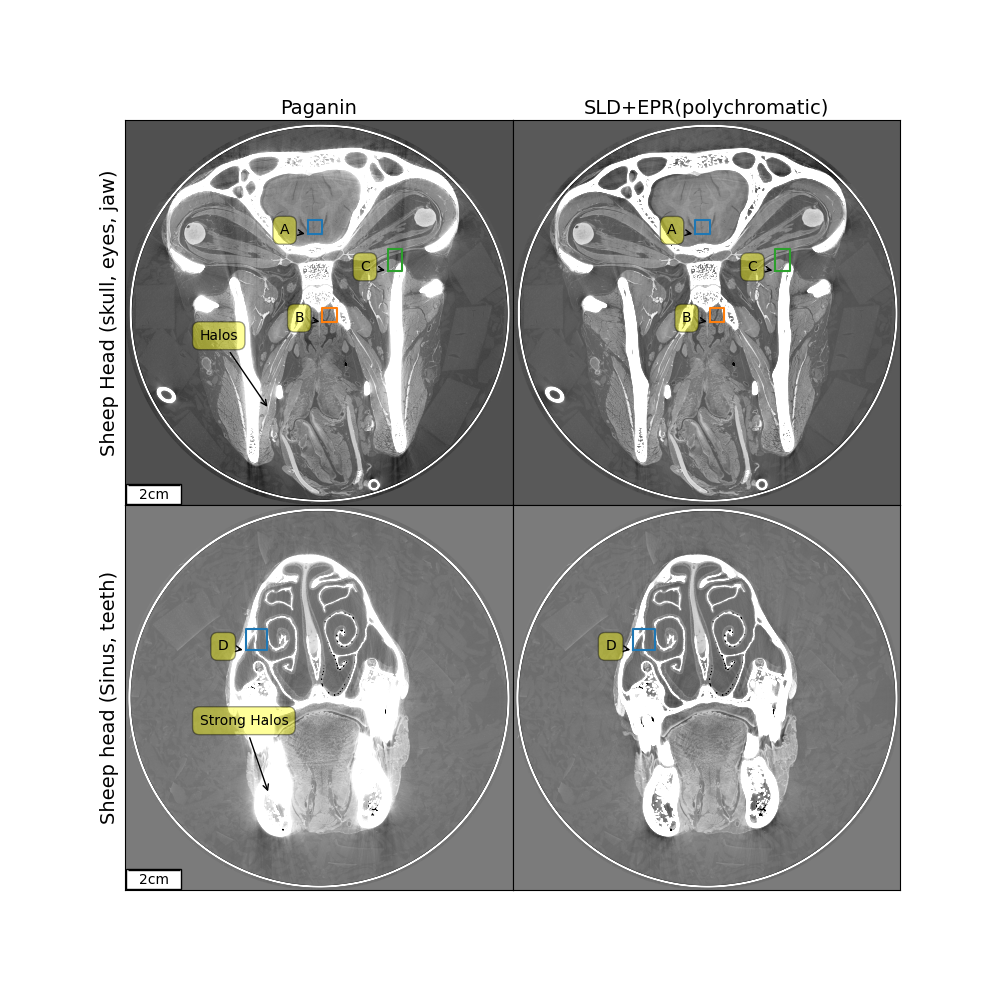}
  \caption{Sheep head HiP-CT scan at average detected energy of 109KeV, 5K x 5K pixels images. Left, the Paganin et al. algorithm at the average detected beam energy. Right, the result obtained using the algorithms introduced by the present work. For EPR we have considered 5 spectral points across the BM18 detected spectrum and we have applied SLD for a tail containing 30\% of the total signal and a damping length of 80 pixels. We used the refractive indexes for 75\% hydroxyapatite 25\% collagen.  The regions marked A, B, C and D are shown in details in Figure 3, highlighting a detailed comparison between EPR and Paganin et al. methods.}
 \label{sheephead}
\end{figure}

The spectra are calculated \cite{modelisationspectra} given the experimental parameters of which a comprehensive view is provided in Table S1 of the supporting material. The vertical marks, in Figure  \ref{spectramouton}, delimit the energy intervals, each interval having the same integrated area, used for the spectral shape discretisation in the polychromatic method discussed in the present work. Figure \ref{sheephead} shows, for two slices of the sheep head, the comparison between the Paganin et al. algorithm \cite{pag2002} coupled with an unsharp mask filter \cite{unsharp} applied to the radiographs, after the phase retrieval, to recover the fine structures (classical method for single distance PPC-$\mu$CT), and the result obtained by applying all the improved methods introduced by the present work. At the scales visible in Figure \ref{sheephead}, two major improvements can be observed: the complete correction of the bright halos
\begin{figure}[htbp]                                                                                                                                                                                       
\centering\includegraphics[width=0.8 \textwidth ]{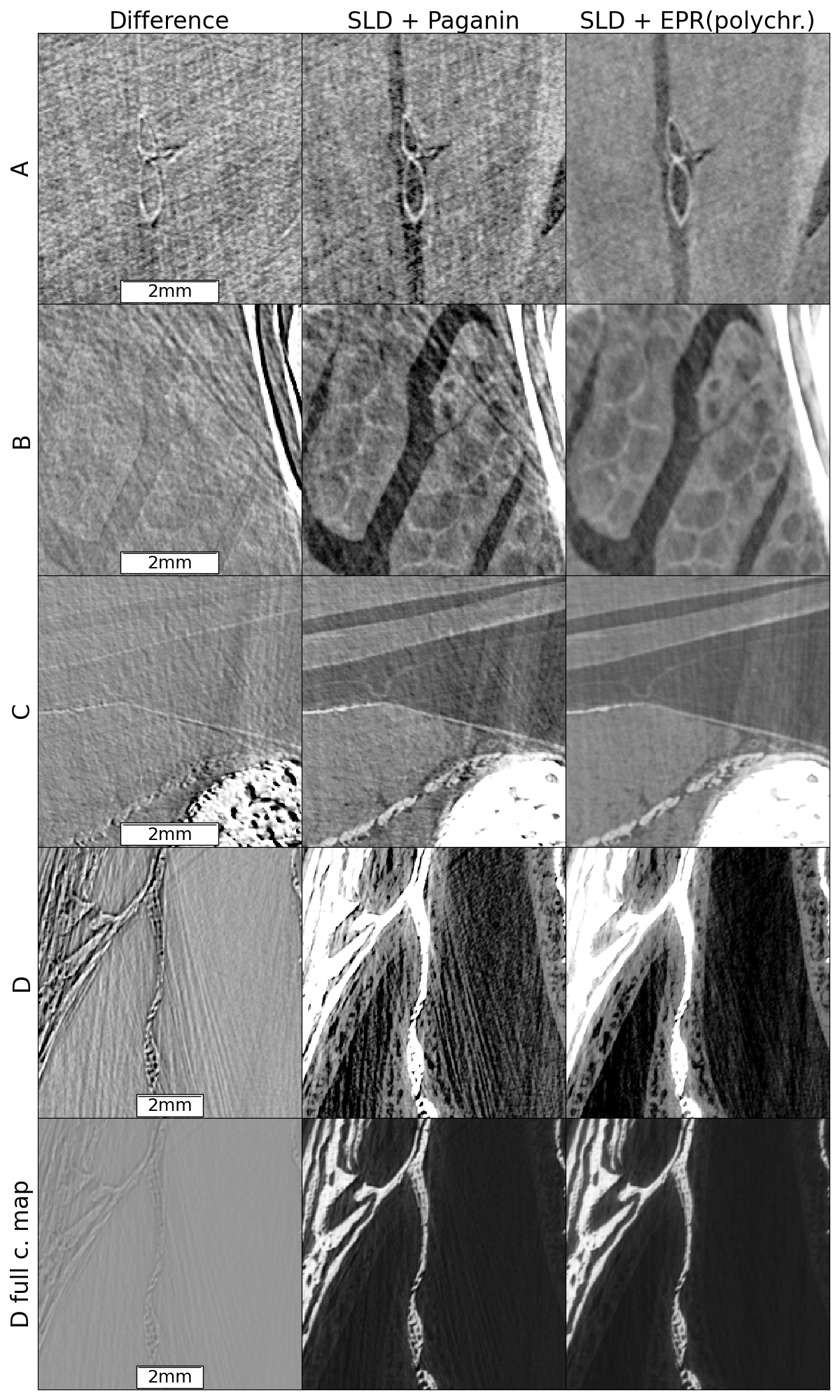}
  \caption{A detailed comparison between the EPR and Paganin et al. methods on zoomed regions from Figure 2 marked A, B, C and D .a detailed comparison between EPR and Paganin et al. method.}
 \label{sheepinsets}
\end{figure}                                                                                                                                          
surrounding the absorbing features and a reduction of the beam-hardening-like artifacts. These improvements are the effect of the SLD method. The improvement brought by the EPR method becomes visible when zooming to smaller length scales. Specifically, the EPR corrects the strong gradient effects which occur on small length scales. For a closer examination of EPR corrections Figure \ref{sheepinsets} provides zoomed views of the insets A (brain), B and C (soft tissues near the bones) and D (nasal mucosa). The gray scales used have a constant min to max width over a given row and are centred, for the four top lines, over the soft matter range. The nasal mucosa is shown twice: first, on the fourth line, shrinking the gray scale window to capture the mucosa. Then using the full range to show the bone details.
Interestingly, the difference image is weakly correlated with the image soft regions features, but strongly with the bone interfaces, especially in tangential directions. These regions typically experience strong gradient, and consequently our algorithm corrects most strongly in these regions. The long range of these artifacts is also noteworthy. Despite being up to 400 pixels from the bones, the errors removed from region A, have a visibility comparable to that of the imaged features. Region C remains the most problematic one, as artifacts corresponding to the long absorption path through elongated bones are not fully corrected. The improvements, highlighted by the region D insets, illustrate the practical benefits of the EPR algorithm. Here a simple threshold-based segmentation method, saturating the bone regions and under-saturating the interstitial ones, if used on the EPR images, is enough to capture the mucosa whose profile would otherwise be lost.

\begin{figure}[htbp]
\centering\includegraphics[width=0.8 \textwidth ]{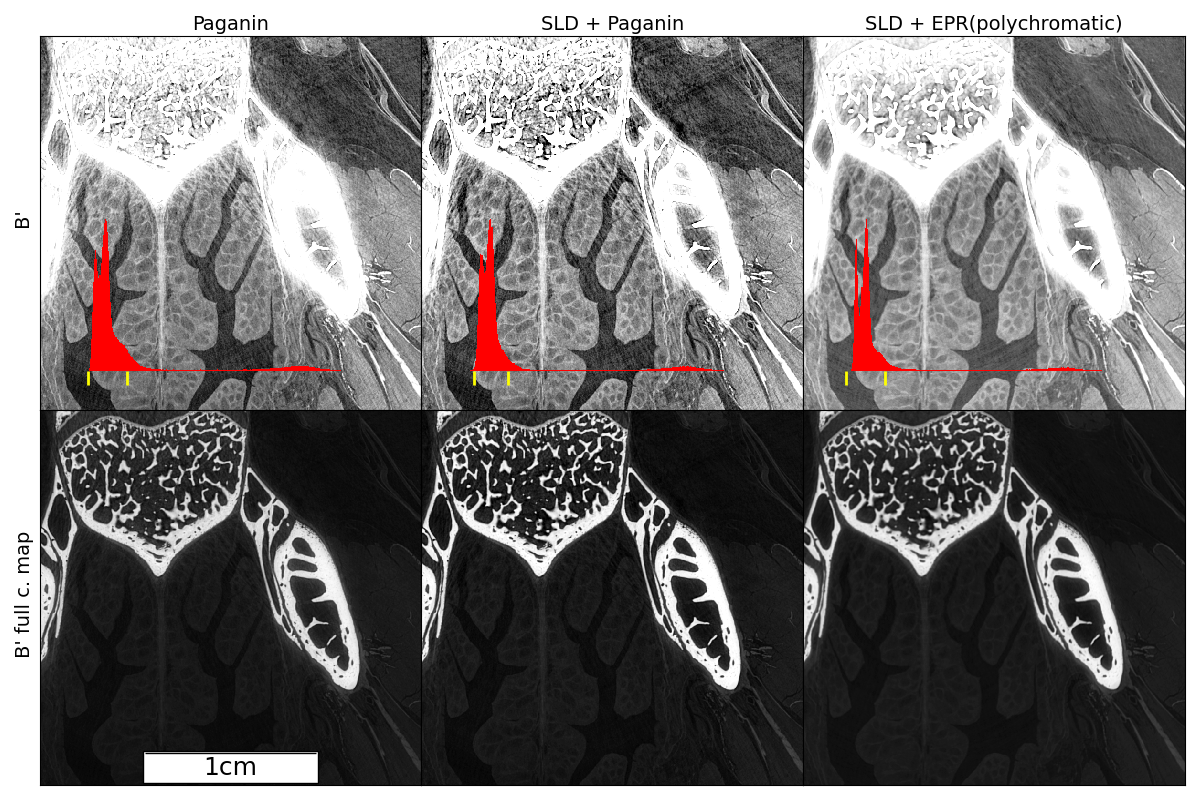}
  \caption{Incremental application of SLD and EPR: (left) reconstruction by the standard Paganin et al. algorithm, (center) the result by preprocessing the radiography with SLD, (right) the effect of both SLD and EPR algorithm. In the first row the gray scale is adapted to the soft matter (the range is marked by yellow ticks in the red histogram), in the second  it covers the whole range. }
 \label{intermediatescalesheep}
\end{figure}

\begin{figure}[htbp]
\centering\includegraphics[width=0.8 \textwidth ]{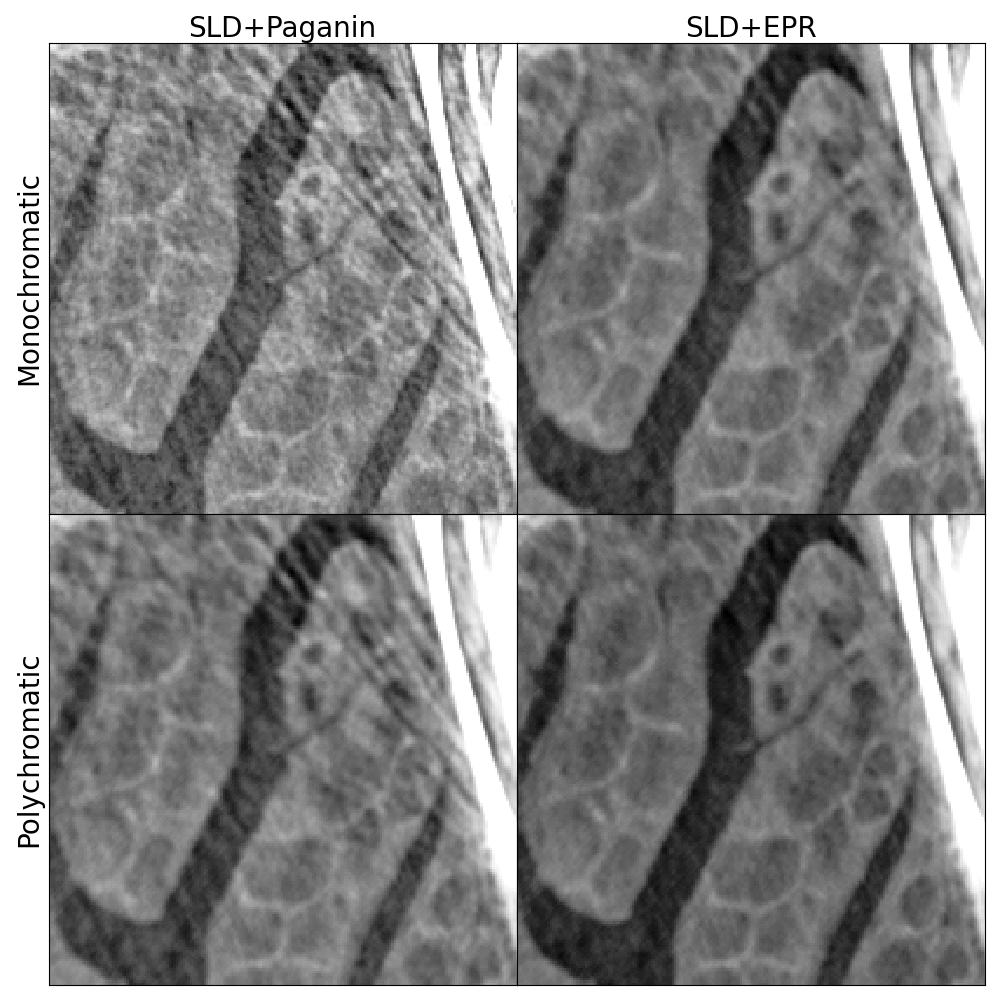}
  \caption{Effect of polychomatic method for EPR and Paganin: first row, monochromatic approach considering only one photon energy  in both the Paganin  and EPR forward model. Second row, polychromatic approach considering the discretised spectra, with five photon energies (as in Figure \ref{spectramouton}), in the forward models. The polychromatic version of the Paganin phase retrieval has been implemented replacing the EPR forward model, for each spectral point, by Eqs. (\ref{linearization},\ref{tie}),  and keeping all the rest of the EPR code unchanged.}
 \label{polyeffect}
\end{figure}

To disentangle the effects of the two algorithms (SLD and EPR) we use an intermediate scale in Figure \ref{intermediatescalesheep} where we show a region B$^\prime$  which is larger than and centered on region B. The figure illustrates the reconstruction by the standard Paganin et al. algorithm (left), the result obtained preprocessing the radiographs with SLD (center), and finally the effect of both SLD and EPR algorithm (right). While, the SLD deconvolution effectively removes the bright halo surrounding the bones, the EPR method's effect (removal of streak artefacts) remains effective and visible even far from the bones which are the source of the artefact. Albeit above evidences, pointing toward the role of the strong gradient and the breaking of the linearization approximation, one might still wonder wether the observed improvement could be ascribed instead to the accounting of polychromaticity done in EPR. To answer this question we show in Figure \ref{polyeffect} a comparison of the Paganin algorithm with EPR for both a monochromatic and polychromatic beam spectral model.

 The polychromatic/monochromatic version of the  Paganin phase retrieval has been implemented replacing the EPR forward model, for a given spectral point, by Eqs. (\ref{linearization},\ref{tie}),  and keeping all the rest of the EPR code unchanged. We can observe that the polychromatic version of Paganin forward model, still produces all the streak artefacts which are instead suppressed at a higher degree by a merely monochromatic EPR algorithm.

\begin{figure}[htbp]                                                                                                                                                                                       
\centering\includegraphics[width=\textwidth ]{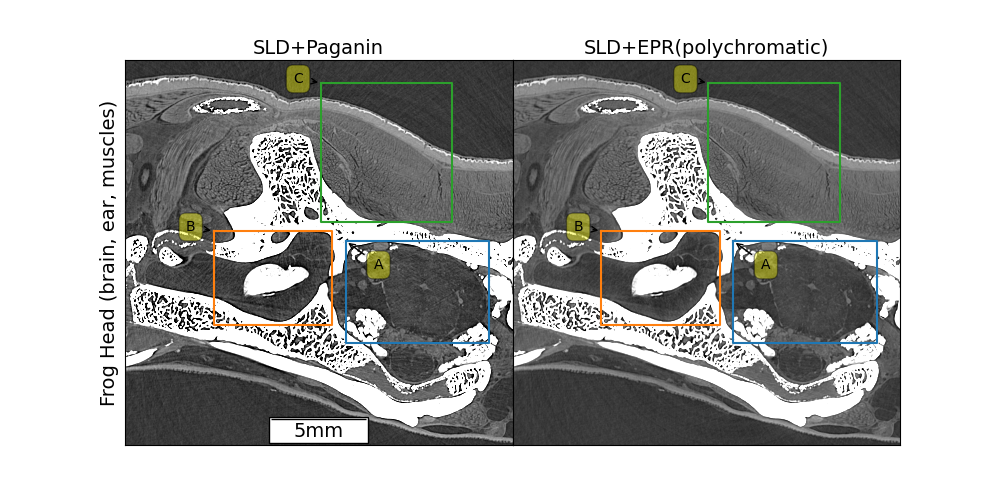}
  \caption{ Frog head: 900 x 800 view spanning the brain (A), ear (B) and skull muscles (C) subregions. The image has been obtained with SLD + Paganin (left) or EPR (right) algorithms. The acquisition has been performed at BM18 with an average beam energy of 127 KeV using the HiP-CT protocol, a voxel size of 23 .27 microns and 20 meters propagation distance.}
 \label{froghead}
\end{figure}

\begin{figure}[htbp]
  \centering\includegraphics[width=\textwidth ]{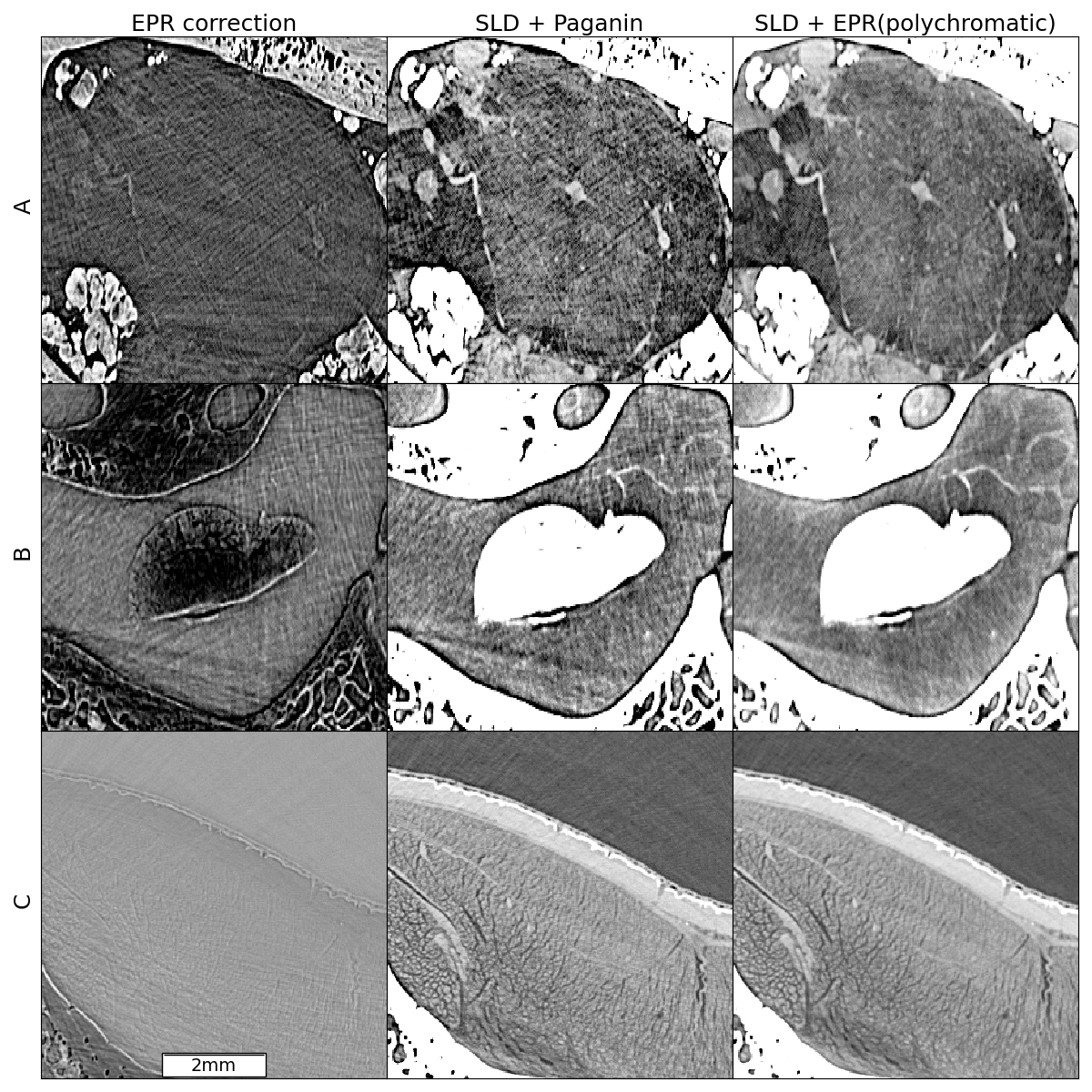}
  \caption{zoomed views on the frog head subregions. Rows: region A, B and C from Figure 5. Columns: left: the difference between the two methods. Center: SLD + Paganin et al.. Right column: SLD + EPR with 5 spectral points. The used grey scales have a constant min to max width over a given row, and are centered, for each image, over the soft matter range.}
  \label{froginsets}
\end{figure}

Figure \ref{froghead} shows a slice subregion from the giant frog head. The acquisition was performed at BM18 with an average beam energy of 127 KeV, a voxel size of 23.27$\mu$m and a propagation distance of 20m using the HiP-CT protocol. The detailed comparison between the classical Paganin et al. approach and EPR is shown in Figure \ref{froginsets} for regions A (cerebral region midbrain), B (membraneous vestibular region of the inner ear (saccular otoconia) and C (dorsalis scapulae muscle). The brain is contained in a mainly convex region and, for this reason, its reconstructed image is not disturbed by those artefacts which are generated at bones borders. They are tangential and do not cross the inner regions. In region A, the Paganin et al. reconstruction artifacts can be attributed to the bone trabecular regions, which, for certain directions, act as an array of refractive lenses. These artifacts are efficiently reduced by EPR. A similar artifact reduction is observed in the ear region. However, the bone tangential artefacts, now visible in this non-convex region, are only weakly reduced by EPR. 
Finally, in the muscles region (row C, Figure \ref{froginsets}), which has a higher signal contrast compared to other soft tissue areas, both Paganin et al. and EPR yield excellent results. For the frog head we have found that the effect of SLD is barely visible. This is due to the relatively small absorption of the skeleton in this sample. The absorption contrast observed in the radiographs between soft and bone regions is in facts always below 15\%, as observed in the intensity modulations of the normalised radiographs. For the sheep head, instead, the contrast peaks at 70\% in certain angles and regions. A practical aspect of the better image quality brought by the EPR+SLD is that automatic segmentation is easier. This is visible already with a simple thresholding method as shown in Figure \ref{sheepinsets}, for region D. In the supplementary material section, the movies M1-M3, which explores the whole 3D volume of the sample, are provided as links to the ESRF data portal.

\section{Material and Methods}

The widely-used retrieval algorithm by Paganin et al. \cite{pag2002} relies on the linearization of the intensity evolution along the propagation axis. To illustrate the limitations of this approximation when strong phase effects occur, we briefly review its framework. 

\subsection{The Paganin et al. Algorithm}

For a point  in the wavefront plane, the intensity at the detector, placed at $z=L$ , is written, for a parallel beam as:

\begin{equation}
I \left( x , y , L \right ) = \left. \frac{  \partial I \left( x , y , z \right)}{\partial z}\right|_{z = 0} L + I \left(x , y , 0 \right) \label{linearization}
\end{equation}

Here the propagated intensity is written, for a point  in the wavefront plan, as the sum of the intensity at  position $(x,y,z=0)$  and a term obtained by multiplying the  propagation distance with the derivative of the intensity respect to $z$ at the sameposition.    
This derivative is obtained, in the following equation, from the transverse derivatives of the intensity and of the phase   by applying the transport-of-intensity equation (TIE) \cite{Teague1983DeterministicPR}:
\begin{equation}
  \frac{\partial I \left(r \right)}{\partial z} = \frac{- 1} {k} { { \nabla} }_{x , y} \cdot \left[ I \left(r \right) {{ \nabla}}_{x , y} \phi \left(r \right) \right] \label{tie}
\end{equation}

Eventually equation \ref{linearization} takes an easily invertible diagonal form after Fourier transformation, assuming the local refracting index ratio   $ \delta {\left( r \right)} / {\beta} \left(r \right)$  remains constant throughout the sample \cite{pag2002}. Then $\phi = \left({\delta} / {\beta} \right) log{\left(I \right)} / {2}$  and the right part of Eq. (\ref{tie}) becomes proportional to the Laplacian of $I$. For the cone beam case, a preliminary transformation to the parallel geometry case must be applied  \cite{Pogany1997}, after which the same formalism applies.
\subsection{A preliminary observation}
Notably, we perform the following observation that will be used to illustrate later the linearization failure: considering a subregion of the $(x,y,z=0)$ plane, with an hypothetical constant intensity $I$ and phase $\phi$, Eq. (\ref{linearization}) propagates its wavefront to the same subregion at $z=L$ with the same value of $I$. This is because, from Eq. (\ref{tie}), $\partial I / \partial z = 0$ for that region.
\subsection{Limitation for strong gradients}
From a geometrical optics perspective, Eq. (\ref{tie}) can be interpreted as the effect of the angular deflection induced by the phase gradient. With respect to the impact point in the absence of the sample, this deflection induces a transverse shift of the photon impact point on the detector plan equal to:
\begin{equation}
  \left({s}_{x} , {s}_{y} \right) = \frac {L} {k} {\nabla}_{x , y} \phi \left(x , y , 0 \right) = \left({\delta} / {\beta} \right) \frac {L} {2 k} \frac{ {\nabla}_{x , y} I}{I} \label{shifts}
\end{equation}

with $2\pi / k =  \lambda$ being the radiation wavelength. The following simple conceptual experiment illustrates the limitation of the linearisation approximation when the phase gradient induced shifts are larger than the size of the sample features. Assume that the intensity is $I = 0$ everywhere in the $z = 0$ plan except within a small circular spot. When the phase gradient induced shifts, at the spot border, are larger than the spot diameter, then photons passing by the region $I(x,y,0) \neq 0$ and with strong gradients, belonging to the spot border, will propagate
to points $(x^\prime, y^\prime, L)$, outside the geometrical projection of the spot, and therefore such that $I(x^\prime, y^\prime,0) = 0$ but because of the deflected photons themselves it must
be $I(x^\prime, y^\prime,L) \neq 0$. These two latter conditions mutually exclude each other in the frame of
Eq. (\ref{linearization}), as in our preliminary observation.

\subsection{The EPR algorithm}

To overcome this issue, we first rewrite the TIE in a form which will inspire us the development of our algorithm. We rewrite Eq. (\ref{tie}) as the continuity equation 
\begin{equation}
   \nabla \left( v_{r} \rho_r\right) = 0 \label{continuity}
\end{equation}

which expresses the conservation law for a gas of particles (the intensity loss due to absorption in air acts as a simple prefactor in paraxial approximation). In our case, the density is given by $\rho_r = I(r)$ and the local average speed by the three components ($\partial_x \phi, \partial_y \phi, \partial_z \phi \simeq 2 \pi/\lambda)$, where the third component is given in paraxial approximation. The equivalence between Eqs. (\ref{tie}) and Eq. (\ref{continuity}) already suggests the idea for a forward model by which we can replace Eq. (\ref{tie}) even in the non-linear case: it would be a kinematic diffusion of particles. Alternatively, we derive an equivalent formulation by writing the Huygens-Fresnel integral for the field at the detector position $(x^\prime, y^\prime)$ in eikonal approximation. The integral is first written in its full form:
\begin{equation}
  I^{1/2}_{\left(x^\prime,y^\prime,L\right)} exp\left(i \phi_{\left(x^\prime, y^\prime\right)}\right) = \int \int
  I^{1/2}_{\left(x,y,0\right)}   exp\left(i (\phi_{(x, y^)}  + \frac{\sum_{\alpha=x,y} \left(\alpha^\prime - \alpha\right)^2  }{2 L}\right)  dx dy 
\end{equation}
and then it is approximated by a sum over eikonal paths, specifically, paths defined by $s_\alpha = \alpha^\prime -\alpha$ with the shits $s_\alpha$ given by  Eq. (\ref{shifts}) for which the first derivatives in $x,y$ of the phase is zero, while  the contributions from the regions where the phase is rapidly varying is neglected. Each eikonal path contributes then with a phase given by its optical path and with an amplitude depending on the second derivatives of the phase. These latter determine how narrow is the region where the phase is slowly varying. The approximation expresses the amplitude $A_{x^\prime y^\prime} = I^{1/2}_{\left(x^\prime,y^\prime,L\right)} exp\left(i \phi_{\left(x^\prime, y^\prime\right)}\right) $, at detector position  $x^\prime,y^\prime$, by the following equation:
\begin{equation}
  A_{x^\prime y^\prime}= \sum_{\partial_{\alpha} \phi L/k = s_\alpha}
  \frac{I^{1/2}_{\left(x,y,0\right)}}{
    \sqrt{
      det\left( H\left(\phi_{xy}\right) L/k + {\mathbf I}\right)}} exp\left(i (\phi_{(x, y^)}  + \frac{\sum\limits_{\alpha=x,y}^{}{ \left(\alpha^\prime - \alpha\right)^2}  }{2 L} \right)  \label{allpaths}
\end{equation}

Here, the summation is performed, for a given detector point , over all the points satisfying Eq. (\ref{shifts}), the eikonal condition. $H$ represents the Hessian matrix, formed by the second derivatives of the phase. Finally, we obtain the forward model, which extends Eq. (\ref{linearization}) to the non-linear regime, by obtaining the intensity  at the detector as the modulus squared of the right member of the above equation, and discarding, in near field approximation, the interference between contributions of different paths:
\begin{equation}
  I_{x^\prime,y^\prime,L} = \sum_{\partial_{\alpha} \phi L/k = \alpha} \frac{I_{x,y,0}}{det\left( H\left(\phi_{xy}\right) L/k + {\mathbf I}\right)}  = \int I_{x,y,0} \prod_{\alpha=x,y}{\delta\left( \alpha - \alpha^\prime + \frac{L}{k} \frac{\partial_\alpha \phi}{\partial \alpha} \right)} dx dy \label{forward}
\end{equation}
The algorithmic implementation of the forward model is dictated by the right member of the equation. The photons passing through each surface element $dxdy$, of the $z=0$ wavefront, are collected at the pixel position $(x+s_x,y+s_y)$.
The justification for discarding the interference terms is based on the short longitudinal coherence length of the pink beam. In near field approximation, $d^2/M$ is larger than the typical wavelength if $d$ is larger than the pixel size. When we consider the shifts due to strong phase gradients, the interference terms, resulting from overlapping wavefront portions, likely originates from paths whose lengths differences are larger than the wavelength. For the weak gradient regimes, where the Paganin et al. approach is still valid, two or more eikonal paths cannot interfere because this would require shifts larger than one pixel. The beam spectral width is therefore not a concern in this case. 
The forward model is built on GPU and gives the simulated intensity at the detector as a functional of an effective absorption distribution. The code may be operated in both polychromatic and monochromatic mode. The latter being a particular case of the former. The EPR algorithm considers a discretisation of the beam spectral distribution. At each discrete energy, the refractive index is calculated from the Chantler tables \cite{xrt,chantler}. In monochromatic mode, and with an uniform unit illumination of the sample, the transmitted intensity that exits the sample is function of the effective absorption $a(x,y)$ that we take as the free variable of the forward model, which now takes in input $I_{ \left (x , y , z = 0 \right)} = exp \left( -  a_{\left(x , y \right)} \right)$  , while the effective phase, used in
Eq. (\ref{forward}), is deduced as  $\phi \left(x , y \right) = - \left({{\delta}_{av}} / {{\beta}_{av}} \right) a_{\left(x , y \right)} / {2}$. Here the symbol $av$ indicates that the considered optical constants are taken at the effective average beam energy, averaged using the effective spectral shape as weight. These constants are calculated for the material of interest that produces the largest gradients leading to these artifacts, and then we subtract the optical index of the embedding material. As our aim is correcting the streak artefacts originating at the bone interfaces we consider, in the calculation of the  ratio, the bones optical index after subtraction by the alcohol index, the latter being the embedding media.

\subsection{EPR with polychromaticity}

When polychromaticity is considered, the whole spectral range is split into a user-chosen number of subranges, with the $i^{th}$ subrange being characterised by its spectral weight $f_i$. The free variable of the forward model is still $a_{(x,y)}$. The fractional intensity $I_i$ of a given spectral range $i$, at the $z=0$  plan, is derived from the effective beam spectral shape and materials optical constants:
\begin{equation}
  I_i \left (x , y , z = 0 \right) = exp \left(- a_{\left(x , y \right)}   \frac{\omega_i \beta_i}{\omega_{av} \beta_{av}} \right) \label{freevarpoly}
\end{equation}

with $\omega_i$ being the subrange weighted average of the photon energy and $\beta_i$ the optical index imaginary part at $\omega_i$.The phase for subrange $i$ is :
\begin{equation}
  \phi_i\left(x, y\right) = -\frac{1}{2} \frac{\omega_i \delta_i}{\omega_{av} \beta_{av}} a(x,y)
\end{equation}  

The variable $a(x,y)$, function of the wavefront position, is found by the conjugate gradient algorithm, minimizing the $L_2$ norm of the difference between the data and the forward model. The comparison reconstructions, done with the Paganin et al. method, are performed with a delta-beta ratio equal to $ \delta_{av}/\beta_{av}$.

\subsection{The SLD algorithm}

The presence of strong gradients in both the phase and absorption creates another kind of artifacts attributed to the secondary light propagation inside the scintillator and its subsequent scattering. The scintillator, an integral part of the experimental setup, is responsible for converting X-rays into visible light. However, a fraction of this light, instead of traveling straight to the imaging optics, remains trapped due to total internal reflections inside the scintillator plate. It then travels until it is either absorbed or scattered, leading to a diffused background around the bright regions of the radiographs, thus altering the values of neighboring regions. In order to deconvolve this signal, we postulate that the point spread function $p_i(j)$, i.e. the signal recorded by the detection setup at a pixel $j$, for a photon which scintillates at the position corresponding to pixel $i$, is:
\begin{equation}
  p_i(j) = (1 - f) \delta_{ij} + \frac{ g (1 - \delta_{ij}) exp( - r_{ij} \alpha)   }{r_{ij}}
\end{equation}
The case of a perfect scintillator, without diffused light, would correspond to $f,g = 0$. In this case, one recovers the ideal spread function $p_i{j}=\delta_{ij}$. A real, non-ideal, scintillator is characterized by a non-zero $f$. This parameter is the fraction of the total collected light which is found in the tail. The parameter $g$ is set by the condition that the sum of $p_i(j)$ on all the pixels is normalised to one. The functional form of the tail contains the $1/r$ decrease of light intensity in 2D geometry, plus the damping $exp( - r_{ij} \alpha )$ due to the light escaping from the scintillator surfaces or being absorbed. The free parameters are $f$ and $\alpha$. To determine these parameters, with the same experimental setup that is used for the tomography experiment, we form a bright small spot on the scintillator by collimating the X-ray beam with slits. Then the spot halo is fitted with the model. The obtained function is then used to deconvolve the raw data.
Finally, for all the reconstructions shown in this work, an unsharp filter \cite{unsharp} has been applied, for every retrieved absorption radiography, following the hierarchical phase-contrast tomography (HiP-CT) protocol \cite{Walsh2021}, with a sigma of $1.2$ pixels for the blurring gaussian kernel and a multiplicative coefficient equal to 4. The EPR retrieval have been performed considering a beam spectral splitting consisting of five contiguous spectral regions, covering the whole energy range, each having the same integrated area.

\section{Conclusion}

\begin{figure}[htbp]
\centering\includegraphics[width=1 \textwidth ]{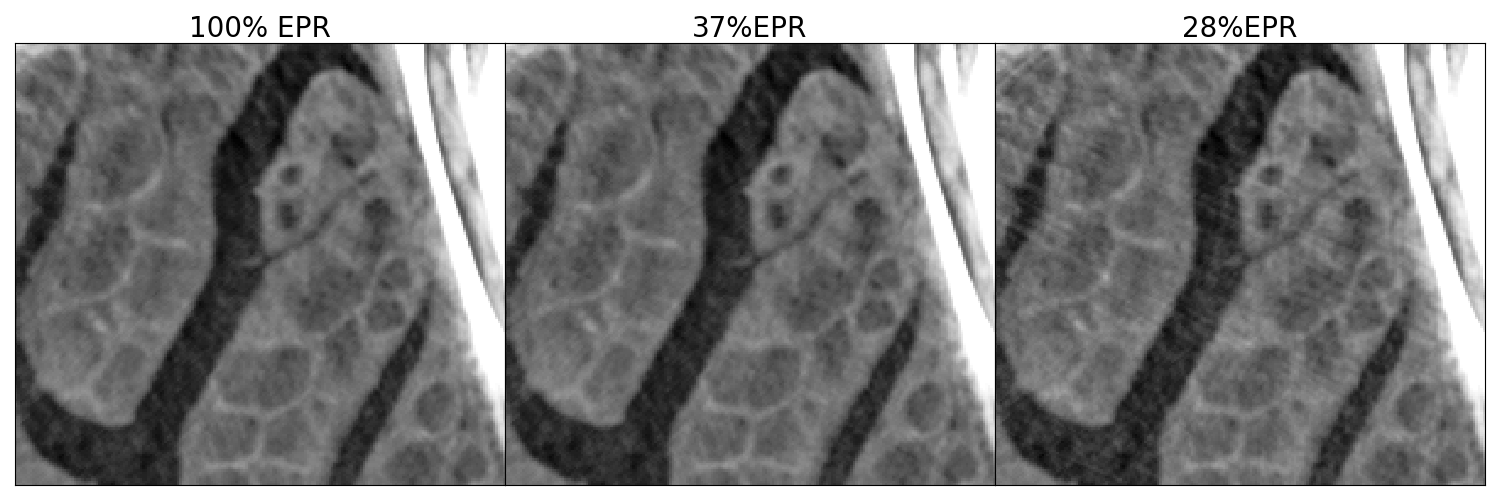}
  \caption{ Hybrid scheme results, where only a fraction of each radiography is processed with EPR and the remaining part with Paganin, according to automatic detection of the crytical regions. From left to right the results are shown for an overall average of 100\%, 37\%, and 27\% of the total projection volume processed with EPR. For an EPR coverage of 37\% the difference with the 100\% case is barely visible. }
 \label{hybrid}
\end{figure}

Our contribution to the image quality improvement is twofold and concerns both low and high frequency artifacts. Concerning the low frequency artifacts, the improvement is based on a simple ansatz for the scintillator response, similar to the approach presented in \cite{deconv}. The strong improvement in the image quality may justify, in the future, more in depth characterization of each scintillator properties, but already our simple implementation gives satisfactory results. Concerning the high frequency artifacts, here lies the most original part of our contribution. We deeply revise one of the most used algorithms in synchrotrons and propose an alternative formulation which copes with large beam deflections. We show drastic suppression of those artifacts which appear as lines originating from strong absorption gradient regions. We stress that our method is based on an improved physical modeling of the beam propagation and simply improves the retrieved signal without artificially post-filtering out any part of it. However, it is worth noting that some residual artifact remains. For instance, in Figure \ref{sheepinsets}, row A and B, the EPR images are still crossed by arrays of residual faint lines, although their visibility has much decreased compared to the classical Paganin et al. images. Moreover, residual artifacts remain strong in regions characterized by long absorption paths running tangentially to large bones thicknesses (Figure  \ref{sheepinsets} row C). This trend is confirmed in the frog sample where the artifacts originating from trabecular structures are strongly suppressed, while the bones tangential artifacts are only weakly reduced. Our hypothesis is that these residual artifacts arise due to the lack of representability of those gradients which occur on slope widths smaller than one pixel. These gradients are underestimated by the discretized numerical procedure which sees them occurring over one pixel. However, despite the residual artifacts, the improvements in image quality are substantial and demonstrate that we are truly leveraging on the inherent physics underlying the observed phenomena. 
The comparison with Paganin et al. algorithm, both monochromatic and polychromatic, highlights the considerable improvement that EPR can bring to the state of the art PPC-$\mu$CT imaging at modern synchrotrons. Beside the aesthetic aspects, our algorithm recovers features, as shown in Figure \ref{sheepinsets} region D, which were lost under the now corrected artifacts. Concerning the residual artifacts, we can base our future strategy, for further improvements, on the logic about gradient representability. Suppose we are interested in a given resolution goal, then the acquisition could be performed at higher resolution. Reducing the pixel size decreases the counts per pixel, with the drawback of increasing the statistical errors. The proper statistics can then be retrieved by transforming the reconstructed image to the lower resolution goal. In fact, it is the signal originating from the soft matter, having the lowest contrast, which is the most affected by the noise. On the other hand, this signal is weak and thus does not create non linearities in the retrieval procedure. The errors will therefore cancel, with random signs, when we revert to lower resolution after the phase retrieval, thus recovering the expected statistics for the goal resolution. The interesting point is that the problematic signal originates from strong gradient regions and these will be better represented, using higher resolution, during the phase retrieval, while their signal is strong and thus more robust to noise. Concerning computing resources, both datasets (for the sheep head and the frog) had a size order of half a terabyte each. When fully processed by the EPR algorithm the computing time was, for one dataset, of the order of one day and a half, using twelve NVIDIA A40 GPUs. In comparison the actual maximum throughput of the beamline is of the order of one similar sample every 4 hours. However not all the measured samples require the EPR algorithm. Moreover, when present, the artifacts have their origin only in particular regions of the radiographs, which are projections, for certain angles of certain strong gradient sample regions. In an ongoing development, the EPR algorithm will be used only in such regions, the Paganin et al., with the SLD, will continue to be used for all the non-critical ones which are the majority of the data, thus reducing the computing time and making the EPR algorithm usable on a routine basis with moderate resources. Figure \ref{hybrid} show the application of this method, obtained by classifying with machine-learning the most-crytical regions of the projections \cite{Leandre}. These preliminary tests show that we can cut, by at least by a factor three, the computing cost.

\section*{Funding.}
\begin{itemize}
\item CZI grant DAF2020-225394 and grant DOI https://doi.org/10.37921/331542rbsqvn from the Chan Zuckerberg Initiative DAF, an advised fund of Silicon Valley Community Foundation (funder DOI 10.13039/100014989).
\item ESRF funding proposals md1252, md1290 and LS-3105.
\item ATTRACT, funded by the EC under Grant Agreement 777222.
\item STAPES ANR-16-CE02-0016.
\end{itemize}
\section*{Acknowledgements.}
Thanks to Michael Krisch for having encouraged, for ATTRACT phase 1 , the submission of the ASPECT project which implemented, in the frame of speckle tomography, a seminal version of the future EPR method, and to Ludovic Broche and Emmanuel Brun for participating in the related experiments. The finalisation of the method and its specialization to PPC-$\mu$CT has been partially financed by the Chan Zuckerberg Initiative. We would like to thank the Royal Belgian Institute of Natural Sciences (IRSNB) and in particular the curator (Recent Vertebrates) Olivier Pauwells for making specimens of the frogs from the Brussels collections available to us.

\section*{Disclosures.}
The authors declare no conflicts of interest.

\section*{Data Availability}
The code sources and a data subset by which the here shown images can be reproduced can be retrieved from the gitlab portal \cite{aspect}, under MIT licence,  and from the ESRF data portal \cite{demodata} respectively.

\bibliography{epr}

\pagebreak

\section{ Supporting Informations}

\subsection{Movie M1.}

Movie M1 referes to File movie\_X3\_combined\_HD.avi

Please find it at https://data.esrf.fr/doi/10.15151/ESRF-DC-1415285771
The video shows a 3D exploration, in parallel for three algorithms, of the sheep head.
      From left to right visible are : 1) Paganin, 2) SLD + Paganin, 3) SLD +EPR.
      The effects of SLD are always visible, around the absorbing features.
      The effects of EPR are visible mainly toward the end of the movie, in the nasal
      mucosa region, and in the brain region in the cerebellum. 

\subsection{Movie M2.}

Movie M2 referes to File movie\_X3\_combined\_HD.mp4

Please find it at https://data.esrf.fr/doi/10.15151/ESRF-DC-1415285771
Video M2 is the compressed version (mp4 format) of video M1.

\subsection{Movie M3.}

Movie M3 refers to Sheep-head\_MRI-HiP-CT\_V2\_100s.mp4

Please find it at https://data.esrf.fr/doi/10.15151/ESRF-DC-1415285771

The video starts with a comparison, for the sheep head, between left)  MRI and right) HiP-CT with SLD and EPR. Then the video first explores volume slice-wise manner and, at about 2/3 of the duration, it starts showing segmented 3d organs.

\subsection{Experimental parameters.}
\begin{center}
\begin{table}
 \caption{Table S1: Overview of the experimental parameters.}
  \begin{tabular}{ |  V{4.5cm} | V{4.5cm} | V{4.5cm} | }
    \hline    
Sample &
Sheep head &
Leptodactylus pentadactylus IRSN-391F \\ \hline

Optic &
Lafip 2 (x0.125) &
dzoom Hasselblad (x0.24) \\ \hline

Average detected energy &
~109 &
~127 \\ \hline

Filters (mm) &
Mo 0.44 &
Mo 1.33 \\ \hline

Water equivalent surface dose rate on sample &
38.4 Gy/s &
14.6 Gy/s \\ \hline

Absorbed dose by the bulk sample &
18.4 kGy &
14.4 kGy \\ \hline

Propagation distance (m) &
30 &
20 \\ \hline

Sensor &
iris15 &
sCMOS PCO edge 4.2 \\ \hline

ROI (HxV) &
3104*256 &
2048*400 \\ \hline

Scintillator &
LuAG:Ce 2mm &
LuAG:Ce 2mm \\ \hline

X-ray source &
BM18 lateral beam (0.85T) &
BM18 lateral beam (0.85T) \\ \hline

machine filling mode &
200 mA &
200 mA \\ \hline

Number of projections &
8000 &
6000 \\ \hline

Scan geometry &
HiP-CT using 140mm diameter jar immersed in 150mm tube, 360\textdegree, half-acquisition 1100 pixels, vertical series 5 mm &
HiP-CT using 100mm diameter jar without immersion, 360\textdegree, half-acquisition 900 pixels, vertical series 7 mm \\ \hline

Subframe time (s) &
0,006 &
0,02 \\ \hline

Exposure time (s) &
0,03 &
0,08 \\ \hline

Accumulation level &
5 &
4 \\ \hline

Time per scan (min) &
4.9 &
9.3 \\ \hline

Number of scans &
44 &
29 \\ \hline

Total time (h) &
3.6 &
4.5 \\ \hline

classical reconstruction protocol & 
HiP-CT normalisation by reference jar, single distance phase retrieval with filtered back-projection, vertical concatenation, 16 bits conversion, ring artefacts correction, jp2 conversion & 
HiP-CT normalisation by reference jar, single distance phase retrieval with filtered back-projection, vertical concatenation, 16 bits conversion, ring artefacts correction, jp2 conversion \\ \hline 
\end{tabular}
\end{table}
\end{center}

\section*{L\'eandre Admans Report (Supplementary Material)}

The complete internship report by L\'eandre Admans, carried out at ENSE3 Grenoble within the framework of this research, is provided as supplementary material. 

\textbf{How to access:} \\
Please download the \textbf{source files} ("Download source" option) from the arXiv page of this submission. The report, named \texttt{2024\_Admans\_BLA\_2.pdf}, is included in the source package, together with the LaTeX files of this manuscript.

\vspace{2mm}
\noindent
\textit{Note: The report is not included in the main PDF, but is available as an additional file in the arXiv source package.}

\end{document}